\documentclass[10pt,twocolumn,oneside,journal]{IEEEtran}
\usepackage[utf8]{inputenc}
\usepackage{amsmath}
\usepackage{amssymb}
\usepackage{abbrevs}
\usepackage{subfig}
\usepackage{hyperref}
\usepackage{graphicx}
\usepackage{setspace}
\usepackage{algorithm, algorithmic}
\usepackage{tikz}
\usepackage{abbrevs}
\usepackage{array}
\usepackage{float}

\usepackage{bm} % Needed to allow for bold Greek letters, e.g. psi

\newcommand{\figref}[1]{Fig.~\ref{#1}}
\newcommand{\tabref}[1]{Tab.~\ref{#1}}
\newcommand{\eqnref}[1]{Eqn.~(\ref{#1})}

\newcommand{\secref}[1]{Section~\ref{#1}}
\newcommand{\algoref}[1]{Algorithm~\ref{#1}}

\newcommand\vb{\ensuremath{\mathbf{b}}}

\newcommand\vg{\ensuremath{\mathbf{g}}}

\newcommand\vw{\ensuremath{\mathbf{w}}}
\newcommand\vx{\ensuremath{\mathbf{x}}}
\newcommand\vy{\ensuremath{\mathbf{y}}}

\newcommand\valpha{\ensuremath{\bm{\alpha}}}

\newcommand\mA{\ensuremath{\mathbf{A}}}

\newcommand\mC{\ensuremath{\mathbf{C}}}
\newcommand\mD{\ensuremath{\mathbf{D}}}

\newcommand\mH{\ensuremath{\mathbf{H}}}
\newcommand\mI{\ensuremath{\mathbf{I}}}

\newcommand\mP{\ensuremath{\mathbf{P}}}

\newcommand\mX{\ensuremath{\mathbf{X}}}

\newcommand\mTheta{\ensuremath{\bm{\Theta}}}
\newcommand\mPsi{\ensuremath{\bm{\Psi}}}

\begin{document}
\singlespacing

\title{Compressive Sensing for Spread Spectrum Receivers}

\author{Karsten Fyhn,~\IEEEmembership{Member,~IEEE,}
        Tobias L. Jensen,~\IEEEmembership{Member,~IEEE,}
        Torben Larsen,~\IEEEmembership{Senior Member,~IEEE,}
        and~S{\o}ren Holdt Jensen,~\IEEEmembership{Senior Member,~IEEE}
	        \thanks{The authors are with Aalborg University, Faculty of Engineering and Science, Department of Electronic Systems, DK-9220 Aalborg, Denmark.
            The authors' e-mails are: \{kfn,tlj,tl,shj\}@es.aau.dk. 
            This work is supported by The Danish Council for Strategic Research under grant number 09-067056.
            }
}

\maketitle
%Alphabetically sorted.

%A:
\newabbrev\AWGN{Additive White Gaussian Noise (AWGN)}[AWGN]
%B:
\newabbrev\BER{Bit Error Rate (BER)}[BER]
%C:
\newabbrev\CDMA{Code Division Multiple Access (CDMA)}[CDMA]
\newabbrev\CS{Compressive Sensing (CS)}[CS]
\newabbrev\CSS{Compressive Spread Spectrum (CSS)}[CSS]
%D:
%E:
\newabbrev\EbNo{energy per bit per noise spectral density (Eb/N0)}[Eb/N0]
%F:
%G:
%H:
%I:
%J:
%L:
%M:
\newabbrev\MSE{Mean Squared Error (MSE)}[MSE]
%N:
%O:
%P:
\newabbrev\PRN{Pseudo-Random Noise (PRN)}[PRN]
%Q:
%R:
\newabbrev\RD{Random Demodulator (RD)}[RD]
%S:
\newabbrev\SNR{Signal to Noise Ratio (SNR)}[SNR]
%T:
%U:
%V:
%W:
%X:
%Y:
%Z:

\begin{abstract}
% Introduction
With the advent of ubiquitous computing there are two design parameters of wireless communication 
devices that become very important: power efficiency and production cost.
Compressive sensing enables the receiver in such devices to sample below the 
Shannon-Nyquist sampling rate, which may lead to a decrease in the two design parameters.
% Method
This paper investigates the use of \CS in a general \CDMA 
receiver. We show that when using spread spectrum codes in the signal domain, the 
\CS measurement matrix may be simplified. This measurement
scheme, named \CSS \hskip-0.35em, allows for a simple, effective receiver design. 
% Results
Furthermore, we numerically evaluate the proposed receiver in terms of bit error rate under
different signal to noise ratio conditions and compare it with other receiver structures. 
These numerical experiments show that though the bit error rate performance is degraded by the
subsampling in the \CS \hskip-0.35em-enabled receivers, this may be remedied by including quantization in the receiver
model. 
% Analysis
We also study the computational complexity of the proposed receiver design under different sparsity and measurement ratios. 
% Discussion
Our work shows that it is possible to subsample a
\CDMA signal using \CSS and that in one example the \CSS receiver outperforms the classical receiver.
\end{abstract}


%%%%%%%%%%%%%%%%%%%%%%
\section{Introduction}
%%%%%%%%%%%%%%%%%%%%%%
% General introduction
As wireless communication devices are becoming more and more widespread and ubiquitous, the need for power efficiency and low production cost becomes paramount.
A power costly operation in wireless communication is the conversion from analog to digital signals - the Analog to Digital Converter (ADC).
The classic ADC uses the Shannon-Nyquist sampling theorem to represent an analog signal in digital form.
The Shannon-Nyquist sampling theorem states that to perfectly represent an analog signal, it must be sampled at a frequency higher than twice the signal's bandwidth.
When this theorem is obeyed, the original analog signal may be reconstructed perfectly from its sampled representation.
The Shannon-Nyquist sampling theorem has been the foundation of digital signal processing for more than half a century and is considered a fundamental building block of digital signal processing systems.
Recently, a new concept termed Compressive Sensing (CS) \cite{Candes2006c,Donoho2006} has been attracting more and more attention in the signal processing community
as it provides an exception to the lower bound on the sampling rate by exploiting sparsity in the signal.
If a signal is sparse in some arbitrary basis, it may be sampled at a rate lower than the Nyquist frequency.
Sparsity in \CS is when a signal is comprised of only a few atoms from a given basis.
The sampled signal must be acquired through some linear measurement scheme.
Examples of these are random Gaussian, Bernoulli and Rademacher measurement schemes, as well as the \RD \cite{Kirolos2006, Tropp2010} and the Modulated Wideband Converter \cite{Mishali2010}.

% Previous work
Compressive sensing has primarily been studied in the general signal processing area, and relatively few researchers have looked into its application in communication systems.
In \cite{Paredes2007a,Zang2009} the authors examine the use of \CS in Ultra-Wideband (UWB) communication systems for channel estimation where the sparsity of the signal lies in the time domain.
Others have used compressive sensing for source coding in communication networks, together with network coding \cite{Feizi2011}.
In the spread spectrum area, some researchers have studied the general use of \CS for spread spectrum communication systems \cite{Aggarwal2009}.
However, their work is mainly focused on using \CS for fast multi-user detection, rather than subsampling.
Another example is in \cite{Li2012}, where the authors use \CS to decrease the sampling rate of a GPS receiver by exploiting sparsity in the number of possible signal components at the receiver.
Their receiver structure is based on possibly complicated hardware filters, which may make their implementation very difficult 
considering the impact of hardware filters to \CS performance \cite{Pankiewicz2011}.
In \cite{Applebaum2012} the authors treat a similar topic where they design spread spectrum codes to enable a base station to perform multi-user detection on a large number of users, of which only a few are active at a time.
Their work focuses on simple on-off signalling, i.e. the existence of nodes, rather than communication with them, and solves the multi-user detection problem using an adapted convex optimization algorithm. 
Their motivation is on increasing the number of active users in a network, rather than decreasing the sampling rate of the ADC.
A more applied approach is taken in \cite{Fyhn2012} where compressive signal processing \cite{Davenport2010} is applied to enable subsampling of an IEEE 802.15.4 Direct Sequence Spread Spectrum (DSSS) signal.
In \cite{Xie2011} the authors also solve a multiuser detection problem using compressive sensing, but in their work the focus is on the design of possibly complex analog filters. For this paper we focus on keeping the analog part as simple as possible and process the signals in the digital domain instead.

% This work
In our work we apply \CS to a general \CDMA system.
We show that a \RD implementation may be used to subsample the \CDMA signal,
but we also develop a simplified version of the \RD which performs equally well for \CDMA signals
but is simpler and cheaper to implement.
Our motivation is that by taking fewer samples we may be able to conserve power in the receiver, 
as can be seen in e.g. Eqn. 13 in \cite{Kenington2000}.
We show the performance of the proposed receiver structure for the simple discrete case, when compared to a classic receiver structure and an \RD receiver structure.
Then we extend our results to a full RF numerical simulation and demonstrate that the performance is identical in this setting.
Due to noise folding the \CS approach suffers a penalty for downsampling,
but we show that if quantization is taken into account \CS outperforms the classic receiver in some cases.
We finally investigate the complexity of the developed algorithms and compare the computational cost of the numerical experiments with the theoretically calculated computation cost.
Following the paradigm of Reproducible Research \cite{Vandewalle2009}, all our results and code are made available at \url{http://www.sparsesampling.com/css}.

% Notation
To define our notation, let all vectors and matrices be denoted using lower and upper case letters in bold, $\vx$ and $\mX$, respectively.
The Penrose-Moore pseudo-inverse is denoted as $\mX^\dagger$, the transpose of a real matrix as $\mX^\mathrm{T}$  and the conjugate transpose of a matrix as $\mX^\ast$.
The expectation operator is denoted by $\mathbb{E}[\cdot]$.

% "Table-of-Contents"
In the following, we first develop a simple signal model in \secref{sec:signal_model}, based on a dictionary of Gold sequences.
We then elaborate on what \CS is and which reconstruction algorithm we use in the numerical experiments in \secref{sec:compressive_sensing}.
Furthermore, we define a novel measurement matrix design for spread spectrum receivers and demonstrate numerically how this measurement matrix performs with a Gold dictionary and the Subspace Pursuit reconstruction algorithm.
This performance is compared to that of a Rademacher measurement matrix and a \RD measurement matrix.
This is followed by \secref{sec:discrete_numerical_experiment}, which includes a simple numerical experiment of the different receiver structures.
In \secref{sec:rf_numerical_experiment} we extend the experiment to a full RF simulation with and without quantization.
We then analyze the computational complexity of the proposed method in \secref{sec:complexity_analysis}, after which we conclude the paper in \secref{sec:conclusion}.

%%%%%%%%%%%%%%%%%%%%%%%%
\section{Signal Model}
\label{sec:signal_model}
%%%%%%%%%%%%%%%%%%%%%%%%
First, we consider a purely discrete model of a spread spectrum communication system.
Uncoded information bits are sent in a slotted fashion, with each slot containing a single \CDMA signal.
The system is assumed to be synchronized, which may be obtained by e.g. having a central node or base station transmit beacons, which signify the beginning and end of slots. 
This is how mobile phone networks and some wireless sensor networks operate.
The receiver is considered non-coherent, as information is encoded in the phase,
but we do assume that there is no carrier frequency offset between the transmitter and receiver oscillators.
This is of course not a realistic assumption but it is done to keep the system simple.
Future work should investigate the impact of oscillator drift on performance.
Each slot contains an independent \CDMA signal and the slots are decoded sequentially and independently of each other.

For one slot, define a discrete QPSK baseband signal, $\vx \in \mathbb{C}^{\mathrm{N}\times 1}$ as:
\begin{align}
    \vx = \mPsi \valpha,
    \label{eqn:x}
\end{align}
where $\mPsi \in \mathcal{S}_{\mPsi} \subset \{\pm 1\}^{\mathrm{N}\times \mathrm{N}}$ is an orthogonal or near-orthogonal dictionary,
containing spreading waveforms for transmission, $\mathcal{S}_{\mPsi}$ is the subset of $\{\pm 1\}^{\mathrm{N}\times \mathrm{N}}$ that contains orthogonal or near-orthogonal dictionaries and $\valpha \in \{\pm 1\pm j, 0\}^{\mathrm{N}\times 1}$ is a sparse vector,
that selects which spreading waveform(s) and what QPSK constellation point(s) to send. $\valpha$ is a vector here because we only process one slot at a time and we assume that within a slot, the signal amplitude for each user is constant. That $\valpha$ is assumed to be sparse is justified in some scenarios, which is demonstrated shortly.

An example of a system using the above signal model could be a wireless sensor network in which one node must gather information from any possible neighbors.
Each node has a unique \CDMA sequence assigned, which it uses to transfer information and each node does not know which neighbors it has, 
but it knows all possible \CDMA sequences.
Note that in this signal model $\valpha$ is defined so that all users have identical amplitude.
This is not realistic as the distance between nodes might vary a lot, resulting in differences between amplitude in the received signal components.
We choose this simplification here but the reconstruction algorithm is not limited by this and also works for sparse vectors with different amplitude components.

In cases where the number of active nodes or users in a network is smaller than the total number of possible users,
the vector $\valpha$ may be assumed sparse, which is the enabling factor for \CS.
Cases such as these arise in e.g. mobile phone networks and wireless sensor networks,
where the number of surrounding nodes \emph{may} be large, but is \emph{often} small.

At the receiver the following signal is observed:
\begin{align}
    \vy = \mTheta \left( \vx + \vw \right) = \mTheta\mPsi\valpha + \mTheta\vw,
    \label{eqn:y}
\end{align}
where $\mTheta$ is a measurement matrix, which we shall treat later, and $\vw\in\mathbb{C}^{\mathrm{N}\times 1}$ is \AWGN.
Notice here that we take into account noise folding as the noise is folded down into the compressed domain together with the signal.
This makes the noise colored and has an impact on the demodulation performance, because each time the sampling rate is reduced by one half, the \SNR is decreased by $3~$dB \cite{Treichler2011, Castro2011}.

%%%%%%%%%%%%%%%%%%%%%%%%%%%%%
\subsection{Spread Spectrum Dictionary of Gold Sequences}
\label{subsec:gold_sequences}
%%%%%%%%%%%%%%%%%%%%%%%%%%%%%
In spread spectrum signals, a possible dictionary $\mPsi$ is a set of Gold sequences, as used in e.g. GPS technology \cite{Misra2010}. A set of Gold sequences is a special dictionary of binary sequences with very low auto and cross-correlation properties \cite{Gold1967}.
To generate a Gold dictionary, two maximum length sequences must be generated by two linear feedback shift registers (LFSR).
A maximum length sequence is often denoted an $m$-sequence (it has $m$ elements), and is a special kind of pseudo-random noise sequence generated by a LFSR, such that it is periodic and produces a sequence of length $2^{m}-1$.
It is called a maximum length sequence as its period is at maximum length. 
The reason for the length being $2^m-1$ rather than $2^m$ is that the state where all cells are zero must be avoided.
To obtain an $m$-sequence, the LFSR must be carefully chosen as there is no algorithm for ensuring maximum length.
However, there are many known LFSR setups for varying choices of $m$.
Furthermore, the two $m$ sequences must be chosen so that their periodic cross-correlation is three-valued and takes on only the values $\{-1, -t, t-2\}$,
where:
\begin{align}
    t=\begin{cases} 2^{(m+1)/2}+1 & \text{ for odd $m$ and} \\ 2^{(m+2)/2}+1 & \text{ for even $m$.} \end{cases}
\end{align}

The set of Gold sequences are then generated using two $m$-sequences: $\vg_1$ and $\vg_2$, both of length $N=2^m-1$.
Each Gold sequence in the set is generated as $\vg_1 \oplus \vg_i$ (exclusive or), where $\vg_i$ is $\vg_2$ cyclically shifted by the parameter $i$.
As $i$ can take on values between $1\leq i \leq 2^m-1$, each shift constitutes a candidate for the set, resulting in a dictionary as follows:
Define a $N\times N$ dictionary of Gold sequences as $\mPsi$, with each column signifying a possible code sequence.

When using such a \CDMA dictionary, the received signal must be sampled at a rate corresponding to the chip rate, where a chip is one entry in the received Gold sequences.
If $\valpha$ is sparse the information rate of the signal is much lower and it may be possible to decrease the sampling rate by using CS.

In this paper, we use three Gold dictionary sizes: $m=5, m=7$ and $m=10$.
The $m$-sequence feedback sets used to generate these are:
{\small
\begin{itemize}
    \item $m=5$:    $X^5 + X^2 + 1$ and     $X^5 + X^4 + X^3 + X^2 + 1$
    \item $m=7$:    $X^7 + X^6 + 1$ and     $X^7 + X^4 + 1$
    \item $m=10$:   $X^{10} + X^3 + 1$ and  $X^{10} + X^9 + X^8 + X^6 + X^3 + X^2 + 1$
\end{itemize}
}
The chosen polynomials may be validated by calculating the auto and cross-correlation of the generated dictionaries and verifying that they follow the structure listed in the above.

%shown in \tabref{tab:gold_sequences}.
%\begin{table}[h]
%\centering
%\setlength{\extrarowheight}{3pt}
%{\small
%\begin{tabular}{|l|l|l|}\hline
%    $\mathbf{m}$ & \textbf{1. Polynomial} & \textbf{2. Polynomial} \\\hline
%    $5$     & $X^5 + X^2 + 1$       & $X^5 + X^4 + X^3 + X^2 + 1$\\\hline
%    $7$     & $X^7 + X^6 + 1$       & $X^7 + X^4 + 1$\\\hline
%    $10$    & $X^{10} + X^3 + 1$    & $X^{10} + X^9 + X^8 + X^6 + X^3 + X^2 + 1$\\\hline 
%\end{tabular}
%}
%\caption{$m$-sequence feedback sets in polynomial form used to generate Gold dictionaries of size $m$.}
%\label{tab:gold_sequences}
%\end{table}

%%%%%%%%%%%%%%%%%%%%%%%%%%%%%%%%%%
\section{Compressive Sensing}
\label{sec:compressive_sensing}
%%%%%%%%%%%%%%%%%%%%%%%%%%%%%%%%%%
\CS is a novel sampling scheme, developed to lower the number of samples required to obtain some desired signal.
At the heart of \CS is the linear sampling scheme, called the measurement matrix.
In classic receivers the measurement matrix $\mTheta$ may be modelled as the identity matrix, such that $\vx$ is sampled at the chip rate of each channel ($I$ and $Q$).
Here, we shall denote a classic receiver using $\mTheta_1 = \mI$,
where the subscript $1$ denotes no subsampling and $\mI$ is the identity matrix of size $\mathrm{N} \times \mathrm{N}$.
In \CS another measurement matrix is used.
Denote by $\mTheta_{\kappa} \in \mathbb{R}^{\mathrm{M}\times \mathrm{N}}$ a \CS measurement matrix,
where $\kappa\in\mathbb{N}_1$ is the subsampling ratio when compared to the Nyquist rate and 
$M=N/\kappa$ (If $\kappa$ does not divide $N$, $M$ is rounded to the nearest integer).
This measurement matrix is then responsible for mapping the $N$-dimensional signal $\vx$ to a $M$-dimensional signal $\vy$.
Normally this would make it impossible to recover the original signal,
but under the assumption that $\vx$ is sparse in some basis,
it is possible to reconstruct the original signal from the sampled, $M$-dimensional signal $\vy$ \cite{Candes2006c,Donoho2006}.

Notice that we are not interested in the reconstructed signal, $\vx$, but in the sparse vector $\valpha$,
which allows us to demodulate the data in the signal.
We may obtain an estimate of $\valpha$ by reconstructing the signal from $\vy$.
Such a reconstruction may be obtained using e.g. a convex optimization problem solver or a greedy algorithm.
For this work, we choose the greedy algorithm Subspace Pursuit \cite{Dai2009}.
This algorithm is chosen due to its good performance in terms of both reconstruction accuracy and running time, as shown in \secref{subsec:subspace_pursuit}.

Before explaining the reconstruction algorithm, we return to the measurement matrix and introduce a new measurement scheme which is enabled by the use of \CDMA \hskip-0.35em.
This new measurement scheme is easier to implement than the \RD \hskip-0.35em, but performs almost identically for spread spectrum systems.
We call this a Compressive Spread Spectrum (CSS) measurement matrix and explain it further in the following.

%%%%%%%%%%%%%%%%%%%%%%%%%%%%%%%%%%%%%%%%%%%%%%%%%%%%%%%%%%%
\subsection{Compressive Spread Spectrum Measurement Matrix}
\label{subsec:spread_spectrum_measurement_matrix}
%%%%%%%%%%%%%%%%%%%%%%%%%%%%%%%%%%%%%%%%%%%%%%%%%%%%%%%%%%%
In most \CS literature a choice of measurement matrix or structure must be made.
The Bernoulli or Rademacher distributed measurement matrix is often seen in the theoretical literature, but it is not well suited for practical implementation in a wireless receiver.
The Random Demodulator (RD) sampling structure \cite{Kirolos2006, Tropp2010} is one of the most well-known measurement matrix structures developed, which is well suited for practical implementation.
In the \RD a \PRN sequence is mixed with the received signal followed by low-pass filtering.
Because a spread spectrum transmitter has already spread the signal before transmission, we show that the \RD structure can be improved so that the mixing with a \PRN sequence at the receiver may be skipped.
This is similar to what is done in \cite{Fyhn2012} with IEEE 802.15.4 signals, which uses Direct-Sequence-Spread-Spectrum (DSSS) signals.
These can be viewed as a special class of \CDMA signals, which are used to counter interference from blockers in the same frequency band, rather than to distinguish between users or signals.

The proposed measurement matrix may therefore be defined similarly to the definition of the \RD matrix in \cite{Tropp2010}.
In their work, the measurement matrix is based on two matrices, $\mD$ and $\mH$.
First, let $\epsilon_0, \epsilon_1, \ldots, \epsilon_N \in \{\pm1\}$ be the chipping sequence used in the \RD for a signal of length $N$.
The mapping $\vx \rightarrow \mD\vx$ signifies the demodulation mapping with the chipping sequence, where $\mD$ is the diagonal matrix:
\begin{align}
    \mD = \begin{bmatrix} \epsilon_0 & & & \\ & \epsilon_1 & & \\ & & \ddots & \\ & & & \epsilon_N \end{bmatrix}.
\end{align}

Second, the $\mH$ matrix denotes the accumulate-and-dump action performed after mixing.
Let $M$ denote the number of samples taken and assume here that $M$ divides $N$.
Then each sample is the sum of $N/M$ consecutive entries of the demodulated signal.
The matrix performing this sampling action may therefore be defined as an $M\times N$ matrix,
with $N/M$ consecutive unit entries in the $r$th row starting in column $rN/M+1$ for each $r=0,1,\ldots,M-1$.
An example with $M=3$ and $N=6$ is:
\begin{align}
    \mH = \begin{bmatrix}   1 & 1 &   &   &   &  \\ 
                              &   & 1 & 1 &   &  \\ 
                              &   &   &   & 1 & 1 
          \end{bmatrix}.
\end{align}
The \RD is therefore designed to sample an analog signal, so that in a discrete representation this is the equivalent to:
\begin{align}
    \vy = \mH\mD\vx,
\end{align}
where $\vx$ is the Nyquist sampled input signal and $\vy$ is the compressively sampled output signal.

The reason for applying a chipping sequence is to spread the signal across the frequency spectrum,
so that information is aliased down into the lower frequency area, which is left untouched by the low-pass filtering.
In the proposed receiver this mixing is unnecessary because the signal has already been spread at the transmitter.
The proposed receiver may therefore be simplified to:
\begin{align}
    \vy = \mH\vx.
\end{align}
This is significantly simpler to implement in hardware than the RD.
Comparing to the notation introduced for the measurement matrix in \secref{sec:signal_model} we therefore have: $\mTheta_\kappa = \mH$.

To justify the use of no \PRN sequence in the measurement matrix, consider the following. 
The use of a \CDMA dictionary introduces a random-like dictionary matrix, 
which spreads the signal out so that each sample contains a little bit of the original information signal.
This is similar to what the measurement matrix does in CS.
Therefore, the sampling process may be rewritten as:
\begin{align}
    \vy = \mH\vx = \mH\mPsi\valpha = \mTheta\mI\valpha.
\end{align}
Here, the measurement matrix becomes $\mTheta=\mH\mPsi$, i.e. the subsampling matrix and the \CDMA codes.
The dictionary then becomes the identity matrix.
When viewed like this, it is clear that $\mTheta$ and $\mI$ are incoherent as the identity matrix only takes out one element in $\mTheta$.
Another common mathematical tool for verifying the validity of a measurement scheme for compressive sensing is Restricted-Isometry-Property (RIP).
However, the RIP gives a less precise and more conservative boundary between reconstruction success and failure than other bounds, see e.g. the discussions in \cite{Donoho2010,Blanchard2011}. Instead, phase-transition diagrams \cite{Donoho2010} may be used to demonstrate empirically for which levels of sparsity the dictionary and measurement matrix are applicable.
In the following, we first define the Subspace Pursuit algorithm and then use phase-transition diagrams to show 
that the proposed \CSS measurement matrix has transitions that are very close to those of the Rademacher and \RD measurement matrices for dictionary matrices using Gold sequences.

%%%%%%%%%%%%%%%%%%%%%%%%%%%%%%%
\subsection{Subspace Pursuit}
\label{subsec:subspace_pursuit}
%%%%%%%%%%%%%%%%%%%%%%%%%%%%%%%
To reconstruct the signal a reconstruction algorithm must be chosen.
Many different approaches have been developed, but two main classes of reconstruction algorithms are in widespread use: $\ell_1$ minimization and greedy algorithms.
Often, $\ell_1$ minimization provides the best solution, but if the matrices $\mPsi$ and $\mTheta$ are very large, it is much more efficient to use the simpler greedy algorithms \cite{Maleki2010}.
Therefore, we choose to use greedy algorithms in this work.

In \cite{Maleki2010} an extensive numerical comparison between reconstruction algorithms is performed based on phase transition plots.
Their results show that the best performance is attained using $\ell_1$ (at least theoretically).
Second best is the least angle regression (LARS) algorithm.
However, as shown in Table VII in \cite{Maleki2010}, the LARS algorithm is quite slow.
A better choice is a Tuned Two Stage Thresholding algorithm, which has good performance and is very fast.
In \cite{Maleki2010}, two algorithms in particular are mentioned: CoSaMP and the Subspace Pursuit algorithm.
The Subspace Pursuit algorithm from \cite{Dai2009} is shown to perform best of the two.

\begin{figure}
    \centering
    \includegraphics[width=0.44\textwidth]{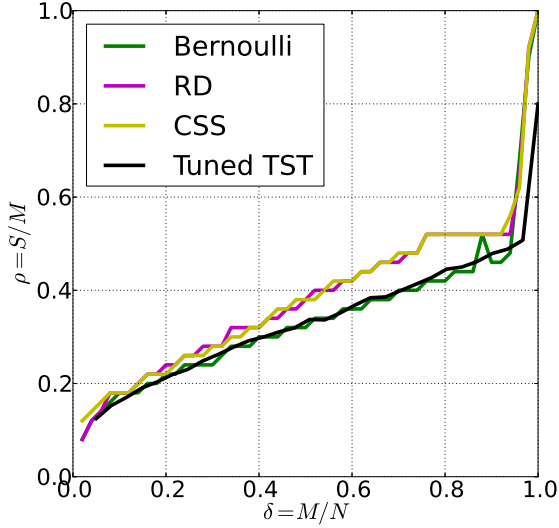}
    \caption{Phase Transition Diagrams for the three different measurement matrices (Rademacher, RD and CSS measurement matrix) with dictionary size $m=10$.
    The black line is the phase transition line for the Tuned Two Stage Thresholding (TST) algorithm from \cite{Maleki2010}.}
    \label{fig:phase_transitions}
\end{figure}

Recall that $\mTheta_\kappa$ is a measurement matrix with $N$ columns and $N/\kappa$ rows and define $\mA = \mTheta_\kappa\mPsi$. Then we define the Subspace Pursuit algorithm as in \algoref{algo:subspace_pursuit}\footnote{In the first initialization step we choose to take the transpose of $\mA$, 
rather than the Penrose--Moore pseudo-inverse.
If instead the Penrose--Moore pseudo-inverse is used, the performance at high values of $\delta$ and $\rho$ is increased in \figref{fig:phase_transitions}, but so is the computational complexity.
This issue is not treated in more detail here, since our problems are assumed to always have low $\rho$.}.
In each algorithm iteration, the pseudo-inverse is calculated as the least-squares solution as this is less computationally demanding.
\begin{algorithm}[h!]
\caption{Subspace Pursuit Algorithm \cite{Dai2009}}
\label{algo:subspace_pursuit}
\begin{algorithmic}
    \STATE \textbf{Input:} 
    \STATE Sparsity $S$, measurement and dictionary matrices combined $\mA$ and received, sampled signal $\vy$
    \STATE \textbf{Initialization:}
        \STATE $T^0 = \{$indices of the $S$ largest absolute magnitude entries \\\hskip1cm in the vector $\mA^\mathrm{T}\vy\}$
        \STATE $\vy^0_r = \vy - \mA_{T^0}\mA_{T^0}^\mathrm{T} \vy$
        \STATE $\ell=0$
    \REPEAT
	    \STATE $\ell \leftarrow \ell+1$
        \STATE $\tilde{T}^\ell \leftarrow T^{\ell-1} \cup \{$indices of the $S$ largest absolute magnitude \\\hskip1cm  entries in the vector $\mA^\mathrm{T}\vy_r^{\ell-1}\}$
        \STATE $T^\ell \leftarrow \{$indices of the $S$ largest absolute magnitude \\\hskip1cm entries in the vector $\mA^\dagger_{\tilde{T}^\ell}\vy\}$
        \STATE $\vy_r^\ell \leftarrow \vy - \mA_{T^\ell}\mA_{T^\ell}^\dagger \vy$
    \UNTIL{$\|\vy_r^\ell\|_2 > \|\vy_r^{\ell-1}\|_2, \ell \geq S$}
\end{algorithmic}
\end{algorithm}

To demonstrate the performance of the Subspace Pursuit algorithm with the Gold dictionary,
we have performed numerical experiments to find the phase transition in the noise-less case for various choices of measurement matrices.
The size of Gold dictionary used is $m=10$, i.e. the dictionary matrix $\mPsi$ is of size $1023 \times 1023$.
The results are shown in \figref{fig:phase_transitions}.
% This figure belongs in the next section, but is placed here as I cannot get LaTeX to place it on the correct page otherwise. :(
\begin{figure*}[ht!]
\centering
\begin{tikzpicture}
    [   state/.style={rectangle,draw, minimum width=2.2cm, minimum height=1cm, text width=1.5cm, fill=gray!30, rounded corners},
        arrow/.style={<-, thick},
        arrow2/.style={->, thick}
    ]
    \node (start) at (0,0) {};
    \node (encoding) at (2,0) [state]                                   {Encoding}
        edge [arrow] node[auto,swap] {$\vb$} (start);
    \node (CDMA mapping) at (5,0) [state]                               {CDMA mapping}
        edge [arrow] node[auto,swap] {$\valpha$} (encoding);
    \draw [-, thick] (CDMA mapping) to node [auto] {$\vx$} (13.5,0);
    \node (noise) at (13.5,-1) [circle, draw]                           {$+$};
    \draw [arrow2] (14.5,-1) to node [auto,swap] {$\vw$} (noise);
    \draw [arrow2] (13.5,0) -- (noise);
    \draw [-, thick] (noise) -- (13.5,-2);
    \node (sampling) at (11,-2) [state]                                 {Sampling};
    \draw [arrow2] (13.5,-2) to node [auto] {$\vx + \vw$} (sampling);
    \node (prewhiten) at (8,-2) [state]                                 {Prewhite}
        edge [arrow] node[auto,swap] {$\vy$} (sampling);
    \node (reconstruct) at (5,-2) [state]                               {Subspace Pursuit}
        edge [arrow] node[auto,swap] {$\mP\vy$} (prewhiten);
    \node (decoding) at (2,-2) [state]                                  {Decoding}
        edge [arrow] node[auto,swap] {$\hat{\valpha}$} (reconstruct);
    \node (end) at (0,-2) {}
        edge [arrow] node[auto,swap] {$\hat{\vb}$} (decoding);
\end{tikzpicture}
\caption{Flow chart of the discrete numerical experiment.}
\label{fig:discrete_numerical_experiment}
\end{figure*}
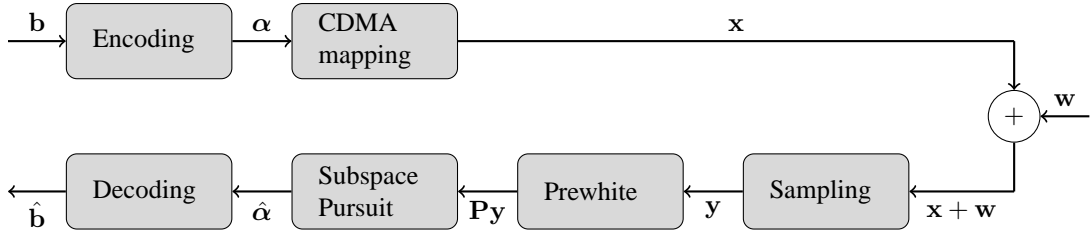
For each curve, we generate a surface plot of the rate of success, based on Monte Carlo simulations.
In this surface plot, a clear transition curve is evident and to condense the results we only plot the transition curve where the probability of error crosses $0.5$.
Each surface plot is generated so that new simulations are conducted until the \MSE between the $i$th and the $(i-1)$th figure is less than $10^{-5}$.
For each parameter set and in each simulation, an experiment is a success ($1$) if the \MSE between the reconstructed and the received signal is less than $10^{-6}$ and a failure ($0$) otherwise.
The three measurement matrices used are as follows:
\begin{itemize}
    \item A Rademacher distributed measurement matrix, with a dense structure where entries are either $-1$ or $1$,
    \item A \RD measurement matrix, with a banded structure, where entries are either $-1$ or $1$ on the band and $0$ outside, and
    \item the proposed \CSS measurement matrix.
\end{itemize}

To validate the above results, we have inserted the phase transition line for the Tuned Two Stage Threshold (TST) algorithm from \cite{Maleki2010} in \figref{fig:phase_transitions}\footnote{Data from \url{http://sparselab.stanford.edu/OptimalTuning/main.htm}}.
As can be seen our implementation corresponds well with their results and it is clear that the proposed \CSS measurement matrix performs close to identically to the \RD measurement matrix and that, as previously argued, the $\mD$ matrix is unnecessary.
Notice also the clear horizontal line in the graph around $\delta=0.9$ and $\rho=0.5$.
We analyze this irregularity more in \secref{sec:complexity_analysis}.

%%%%%%%%%%%%%%%%%%%%%%%%%%%%%%%%%%%%%%%%%%%
\section{Discrete Numerical Experiment}
\label{sec:discrete_numerical_experiment}
%%%%%%%%%%%%%%%%%%%%%%%%%%%%%%%%%%%%%%%%%%%
In the above analysis, we have focused on the noise-less case and have shown that the presented dictionary and measurement matrix setup does enable \CS for certain levels of sparsity.
We therefore now return to the original signal model in \eqnref{eqn:y} and investigate the noisy case by carrying out \BER experiments.
In \figref{fig:discrete_numerical_experiment} a flow chart of the numerical experiment is shown.
First, we encode a randomly generated bit sequence $\vb$ to form the sparse vector $\valpha$ from \eqnref{eqn:x}.
The non-zero positions are chosen randomly from a uniform distribution.
Each non-zero position contains a QPSK symbol.
Then, $\valpha$ is used to create a \CDMA signal using the Gold dictionary as $\vx = \mPsi\valpha$.
This signal is then corrupted by additive white Gaussian noise, generated according to a chosen \SNR value.
Here, \SNR is defined as follows:
\begin{align}
    \text{SNR} = \mathbb{E}\left[\frac{\|\vx\|_2^2}{\|\vw\|_2^2}\right] = \frac{\|\vx\|_2^2}{N\sigma^2},
\end{align}
where $w \sim \mathcal{N}(0,\sigma^2 I)$ with $\sigma^2$ the variance of the noise.

At the receiver, the sampling is modelled as in \eqnref{eqn:y} with multiplication by a measurement matrix.
In the simulations we use $\kappa=2$ or $\kappa=4$. As is shown in the phase transition plots previously, the method also works for other choices of $\kappa$ in the noise-less case. However, to clearly demonstrate that our implementation produces the expected $3$~dB drop in performance per doubling of $\kappa$ due to noise folding, we have chosen these two values.
A measurement matrix based on samples obtained from a Rademacher distribution introduces colored noise.
This decreases the performance, unless the signal is prewhitened before the reconstruction algorithm.
This coloring occurs because the rows in the Rademacher matrix are not orthogonal.
In the \RD and \CSS measurement matrices the rows are orthogonal and prewhitening is therefore unnecessary.
The prewhitening is achieved by multiplying the received $\vy$ vector with a new matrix $\mP$ to obtain $\tilde{\vy}=\mP\vy$.
By setting $\mP=\mC^{-1}$, where $\mC$ is e.g. the Cholesky factorization ($\mC\mC^T = \mTheta_\kappa\mTheta_\kappa^T$), the variance of the noise term $\tilde{\vw}=\mP\mTheta_\kappa\vw$ from \eqnref{eqn:y} becomes: 
\begin{align}
    \mathbb{E}[\mP\mTheta_\kappa\vw\vw^T\mTheta_\kappa^T\mP^T] = \sigma^2 \mC^{-1} \mC \mC^T (\mC^{-1})^T = \sigma^2 \mI.
\end{align}

After prewhitening, we reconstruct the sparse vector $\hat{\valpha}$ using the Subspace Pursuit algorithm,
which now also must include the $\mP$ matrix, i.e. $\mA=\mP\mTheta_\kappa\mPsi$.
It is clear that this extra step increases complexity, but note that this step is only performed for the Rademacher measurement matrix.
The $\mA$ matrix must be generated anew for each slot because a new measurement matrix $\mTheta_\kappa$ is generated.
The \RD and \CSS measurement matrices skip this step as their rows are orthogonal.
After obtaining the sparse vector $\hat{\valpha}$, we are able to decode the original bit sequence, $\hat{\vb}$.
\begin{figure*}[ht!]
    \centering
    \subfloat[$m=5$, $\kappa=2$.]{\includegraphics[width=0.5\textwidth]{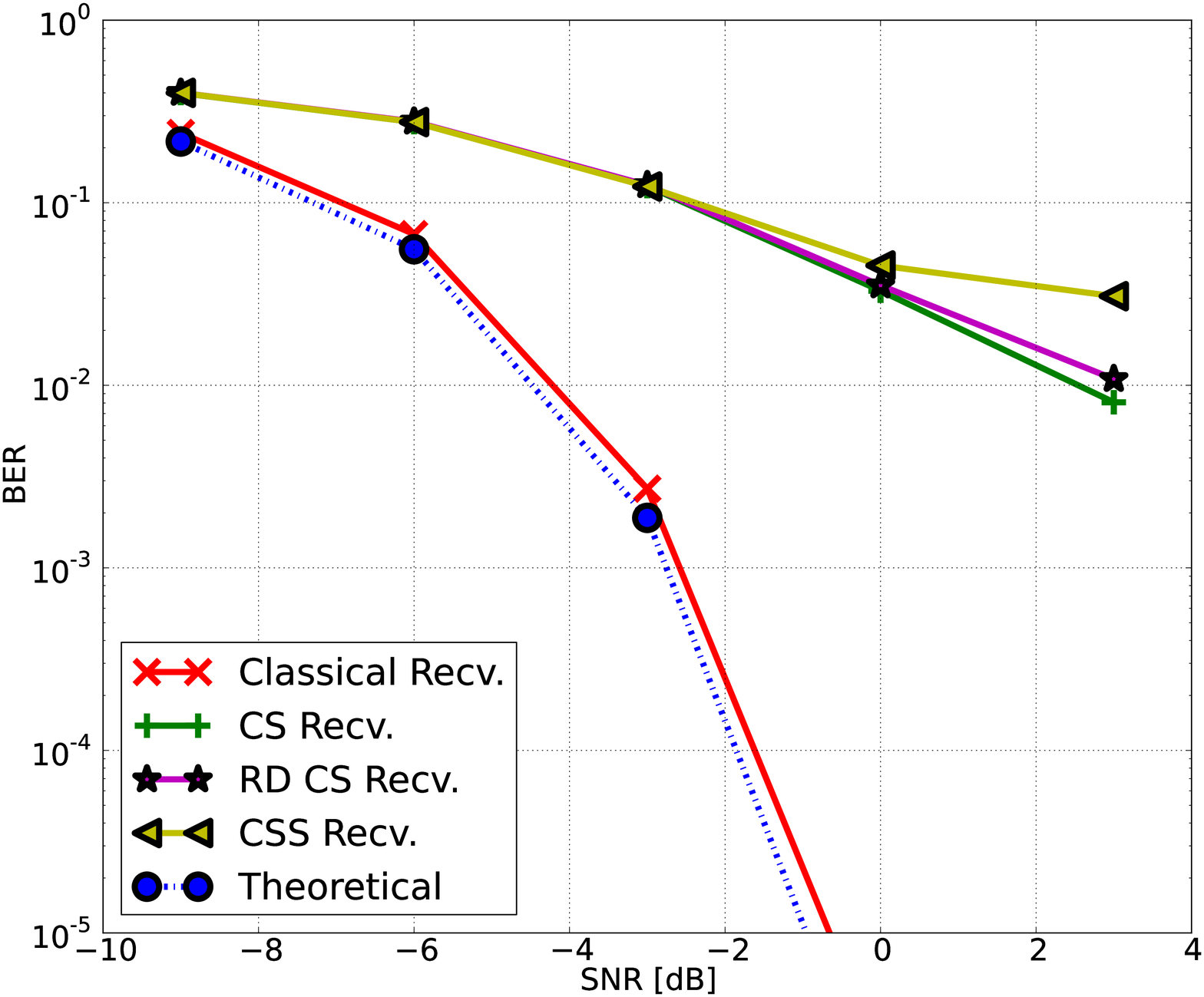}}
    \subfloat[$m=10$, solid lines are for $\kappa=2$, dashed are $\kappa=4$.]{\includegraphics[width=0.5\textwidth]{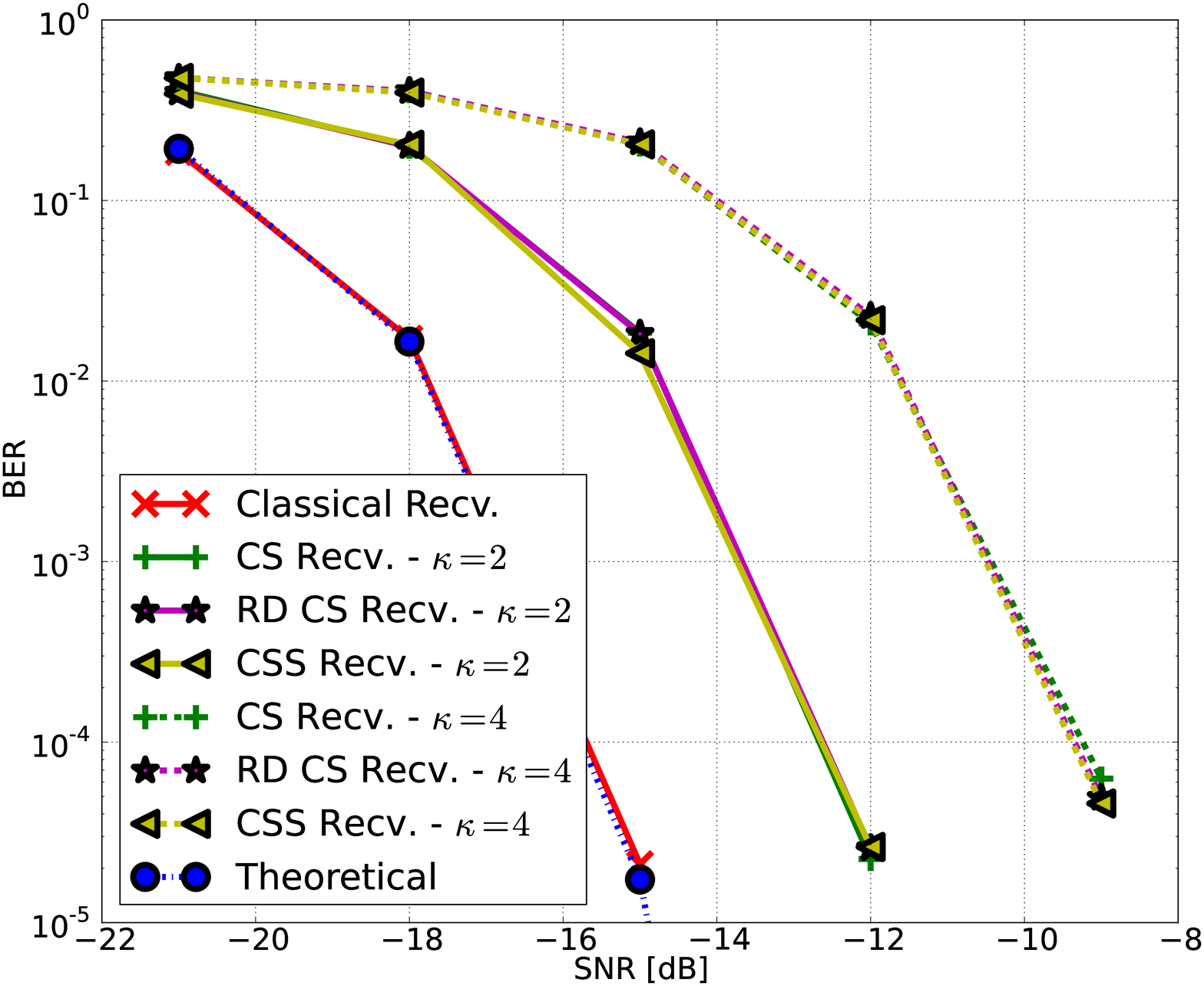}}
    \caption{BER versus SNR for different dictionary sizes and choices of $\kappa$. CS here is the Rademacher measurement scheme. Simulations were run until $100$ bit errors were found for each SNR point.}
    \label{fig:discrete_numerical_experiments}
\end{figure*}

To validate the obtained results, we compare the numerical results with the theoretical performance for non-coherent MFSK \cite{Sklar2001}:
\begin{align}
    \nonumber P_b = \frac{N}{2(N-1)} \frac{1}{N} \sum_{k=2}^N (-1)^k {N \choose k} \cdot \\ \exp\left(N \cdot \text{SNR}\left(\frac{1}{k}-1\right)\right),
\end{align}
where $\text{SNR}$ is the signal to noise ratio.
We use the non-coherent formula because the \CDMA codes are QPSK modulated.
This corresponds to a phase shift of the original codes,
which makes the receiver non-coherent.
Furthermore, for the above result to hold, we must fix $S=1$, i.e. the \CDMA signal is $1$-sparse.
Then, instead of performing reconstruction of the sparse $\valpha$, we may instead perform classification as in \cite{Fyhn2012,Davenport2010}.
This would replace the Subspace Pursuit algorithm with a simpler estimation framework.
However, to conserve generality and because we use $S=10$ later, we continue using the general \CS framework and the Subspace Pursuit algorithm.

As dictionary we use Gold sequences with $m\in\{5, 10\}$.
This reveals the performance for different dictionary sizes and especially $m=10$ is interesting as it is the LFSR length used in e.g. GPS.
The results of the numerical experiments are shown in \figref{fig:discrete_numerical_experiments}.

As can be seen, for $m=5$ both the Rademacher, \RD and especially the \CSS measurement matrix seems to perform poorly.
For high \SNR values there is more than the expected $3$~dB loss per octave due to noise folding. 
At $m=10$ the loss is almost exactly $3$~dB per halving of the sampling rate.
For $m=10$ we have also included the result for $\kappa=4$ to show the performance when the sampling rate is reduced to a quarter of the Nyquist sampling rate.
Again, the curve follows the previous results for noise folding, as the performance degrades by approximately $3$~dB more for all the CS-enabled receiver structures.
These results show that the \CSS measurement matrix, though simpler than all the other measurement matrices, performs equally well in the above experiments for $m=10$.
For small dictionary sizes, its performance is worse.

%%%%%%%%%%%%%%%%%%%%%%%%%%%%%%%%%%%
\section{RF Numerical Experiment}
\label{sec:rf_numerical_experiment}
%%%%%%%%%%%%%%%%%%%%%%%%%%%%%%%%%%%
To obtain more realistic communication-relevant results, we have extended the above discrete numerical experiment to a full transmitter/receiver simulation with RF up and down-conversion and with root raised cosine pulse shaping and matched filter.
This we have done to demonstrate that the results from \figref{fig:discrete_numerical_experiments} translate to a realistic transmitter/receiver system.
The construction of the experiment is visualized in \figref{fig:rf_numerical_experiment}.
This conceptual flow chart also visualizes how the ADC process must be incorporated in a receiver structure to implement the proposed \CSS method.
The experiment we have conducted is based on a QPSK signal with a chip rate of $10^6$ chips per second using a root raised cosine pulse shaping filter with a roll-off factor of $1$. 
This signal is represented in the simulation as sampled at $10$ times that rate, to emulate an analog signal.
The signal is up-converted to an RF frequency of $3$~MHz, i.e. $3$ times the chip rate.
The RF signal is sampled at $12$~MHz, again to emulate an analog signal.
Here, AWGN is added followed by down-conversion again.
The down-conversion is implemented as perfect direct down-conversion.
This is accomplished by first multiplying with a complex exponential, followed by taking an FFT of the signal.
In the output from the FFT, all values above the chip rate are set to $0$, after which the inverse FFT is taken.
At baseband, the sampling is done by a matched filter based on the same root raised cosine that is used for pulse shaping.
The samples are then input to the Subspace Pursuit algorithm, similar to the discrete numerical experiment.
\begin{figure*}[ht]
\centering
\begin{tikzpicture}
    [   state/.style={rectangle,draw, minimum height=1cm, text width=1.55cm, fill=gray!30, rounded corners, text centered},
        shortstate/.style={rectangle,draw, minimum width=1cm, minimum height=0.5cm, fill=gray!30, rounded corners},
        arrow/.style={<-, thick},
        arrow2/.style={->, thick}
    ]
    \node (start) at (0,1.5) {};
    \node (encoding) at (1.5,1.5) [state]                                                       {Encoding}
        edge [arrow] node[auto,swap] {$\vb$} (start);
    \node (CDMA mapping) at (3.7,1.5) [state]                                                   {CDMA mapping}
        edge [arrow] node[auto,swap] {$\valpha$} (encoding);
    \node (Ipulseshape) at (10.5,2.3) [state]                                                   {Pulse shaping};
        \draw [thick] (CDMA mapping) -- (7.9,1.5) -- (7.9,2.3); 
        \draw [thick,->] (7.9,2.3) to node [auto] {$\mathrm{Re}\{\vx\}$} (Ipulseshape);
    \node (Qpulseshape) at (10.5,0.7) [state]                                                   {Pulse shaping};
        \draw [thick] (CDMA mapping) -- (7.9,1.5) -- (7.9,0.7); 
        \draw [thick,->] (7.9,0.7) to node [auto, swap] {$\mathrm{Im}\{\vx\}$} (Qpulseshape);
    \node (Iupconv) at (14,2.3) [circle, draw]                                                  {$\times$}
        edge [arrow] node[auto,swap] {$I_x(t)$} (Ipulseshape);
    \node (Qupconv) at (14,0.7) [circle, draw]                                                  {$\times$}
        edge [arrow] node[auto] {$Q_x(t)$} (Qpulseshape);
    \node (cosup) at (13,1.75)                                                                  {$\cos(\omega_c)$};
        \draw [arrow2] (cosup) -| (Iupconv); 
    \node (sinup) at (13,1.25)                                                                  {$\sin(\omega_c)$};
        \draw [arrow2] (sinup) -| (Qupconv);
    \node (adder) at (15,1.5) [circle, draw]                                                    {$+$};
        \draw [arrow2] (Iupconv) -| (adder);
        \draw [arrow2] (Qupconv) -| (adder);
    
    \node (noise) at (15.5,0) [circle, draw]                                                    {$+$};
        \draw [arrow2] (16.5,0) to node [auto,swap] {$w(t)$} (noise);
        \draw [arrow2] (adder) -| (noise);
    
    \node (Idownconv) at (14,-0.7) [circle, draw]                                               {$\times$};
        \draw [arrow2] (noise) |- (15,-1.5) |- (Idownconv);
    \node (Qdownconv) at (14,-2.3) [circle, draw]                                               {$\times$};
        \draw [arrow2] (noise) |- (15,-1.5) |- (Qdownconv);
    \node (cosdown) at (13,-1.25)                                                               {$\cos(\omega_c)$};
        \draw [arrow2] (cosdown) -| (Idownconv);
    \node (sindown) at (13,-1.75)                                                               {$\sin(\omega_c)$};
        \draw [arrow2] (sindown) -| (Qdownconv);
    \node (Ilpf) at (12.9,-0.7) [shortstate]                                                    {LPF}
        edge [arrow] (Idownconv);
    \node (Qlpf) at (12.9,-2.3) [shortstate]                                                    {LPF}
        edge [arrow] (Qdownconv);
    \node (Imf) at (11.7,-0.7) [shortstate,fill=gray!80]                                        {MF}
        edge [arrow] (Ilpf);
    \node (Qmf) at (11.7,-2.3) [shortstate,fill=gray!80]                                        {MF}
        edge [arrow] (Qlpf);
    \node (Iadc) at (9.6,-0.7) [shortstate,fill=gray!80]                                        {ADC}
        edge [arrow] node[auto] {$I_y(t)$} (Imf);
    \node (Qadc) at (9.6,-2.3) [shortstate,fill=gray!80]                                        {ADC}
        edge [arrow] node[auto,swap] {$Q_y(t)$} (Qmf);
    \node (j) at (8.6, -1.5) {$j$};
    \node (jx) at (8.6,-2.3) [circle, draw, inner sep=0.005cm] {$\times$}
        edge [arrow] node[auto,swap] {} (j)
        edge [arrow] node[auto,swap] {} (Qadc);
    \node (adder2) at (7.9,-1.5) [circle, draw]                                                 {$+$};
        \draw [thick] (Iadc) to node [auto, swap, at end] {$\mathrm{Re}\{\vy\}$} (7.9,-0.7);
        \draw [arrow2] (7.9,-0.7) -- (adder2);
        \draw [thick] (jx) to node [auto, at end] {$j\mathrm{Im}\{\vy\}~$} (7.9,-2.3);
        \draw [arrow2] (7.9,-2.3) -- (adder2);
    \node (prewhiten) at (6.2,-1.5) [state]                                                     {Pre- whitening}
        edge [arrow] node[auto,swap] {$\vy$} (adder2);
    \node (reconstruct) at (3.7,-1.5) [state]                                                   {Subspace Pursuit}
        edge [arrow] node[auto,swap] {$\mP\vy$} (prewhiten);
    \node (decoding) at (1.5,-1.5) [state]                                                      {Decoding}
        edge [arrow] node[auto,swap] {$\hat{\valpha}$} (reconstruct);
    \node (end) at (0,-1.5) {}
        edge [arrow] node[auto,swap] {$\hat{\vb}$} (decoding);
\end{tikzpicture}
\caption{Conceptual flow chart of the RF numerical experiment. 
Note that all continuous variables here are only conceptual. 
In the numerical experiments they are represented as discrete, oversampled sequences. 
Here, MF is a matched filter and LPF is a low--pass filter.
Dark boxes signify components that must be changed compared to a traditional architecture to enable the CS subsampling described in this work.}
\label{fig:rf_numerical_experiment}
\end{figure*}
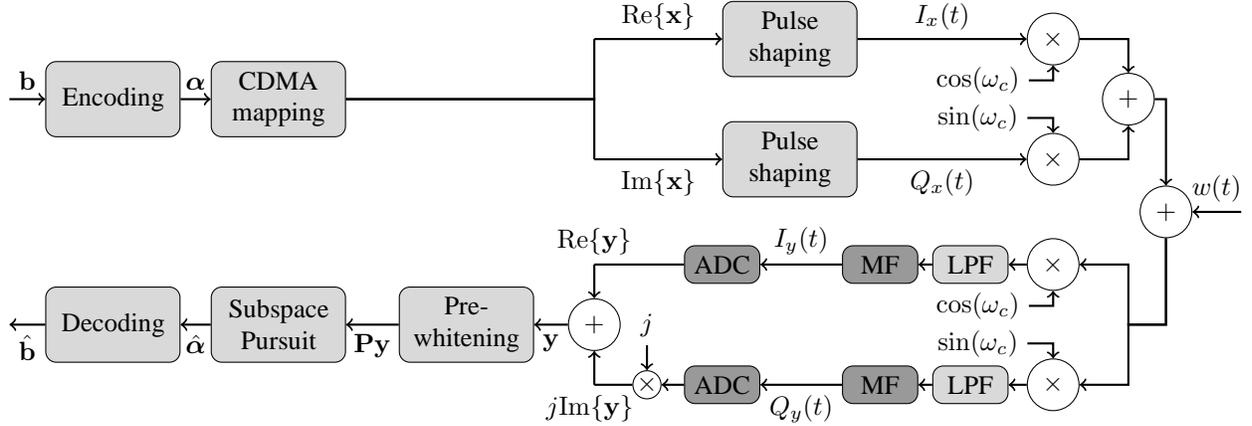
\begin{figure*}[ht!]
    \centering
    \subfloat[$m=5$, $100$ errors found for each point, $\kappa=2$.]{\includegraphics[width=0.45\textwidth]{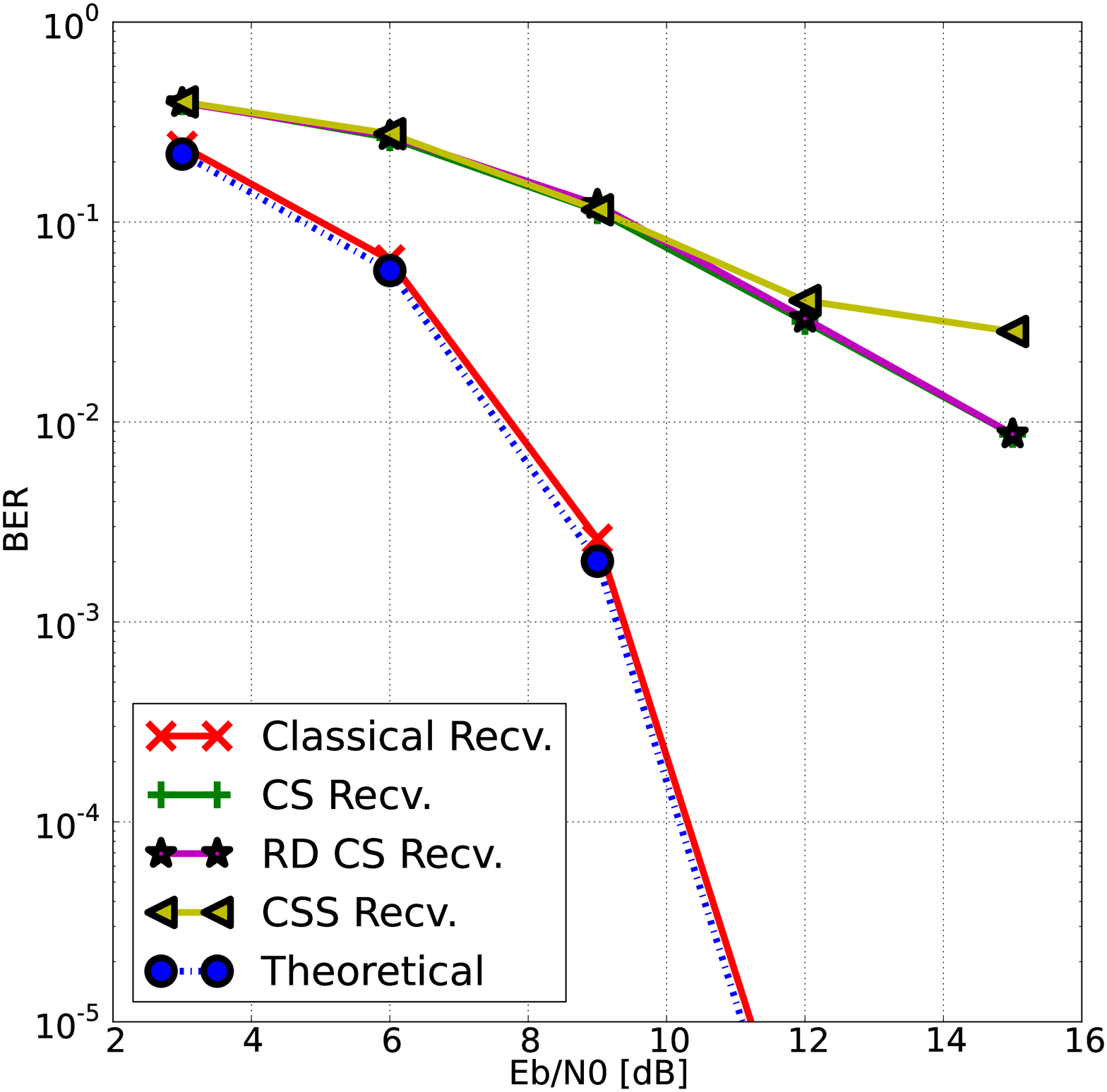}}
    \subfloat[$m=10$, $100$ errors found for each point, $\kappa=2$.]{\includegraphics[width=0.45\textwidth]{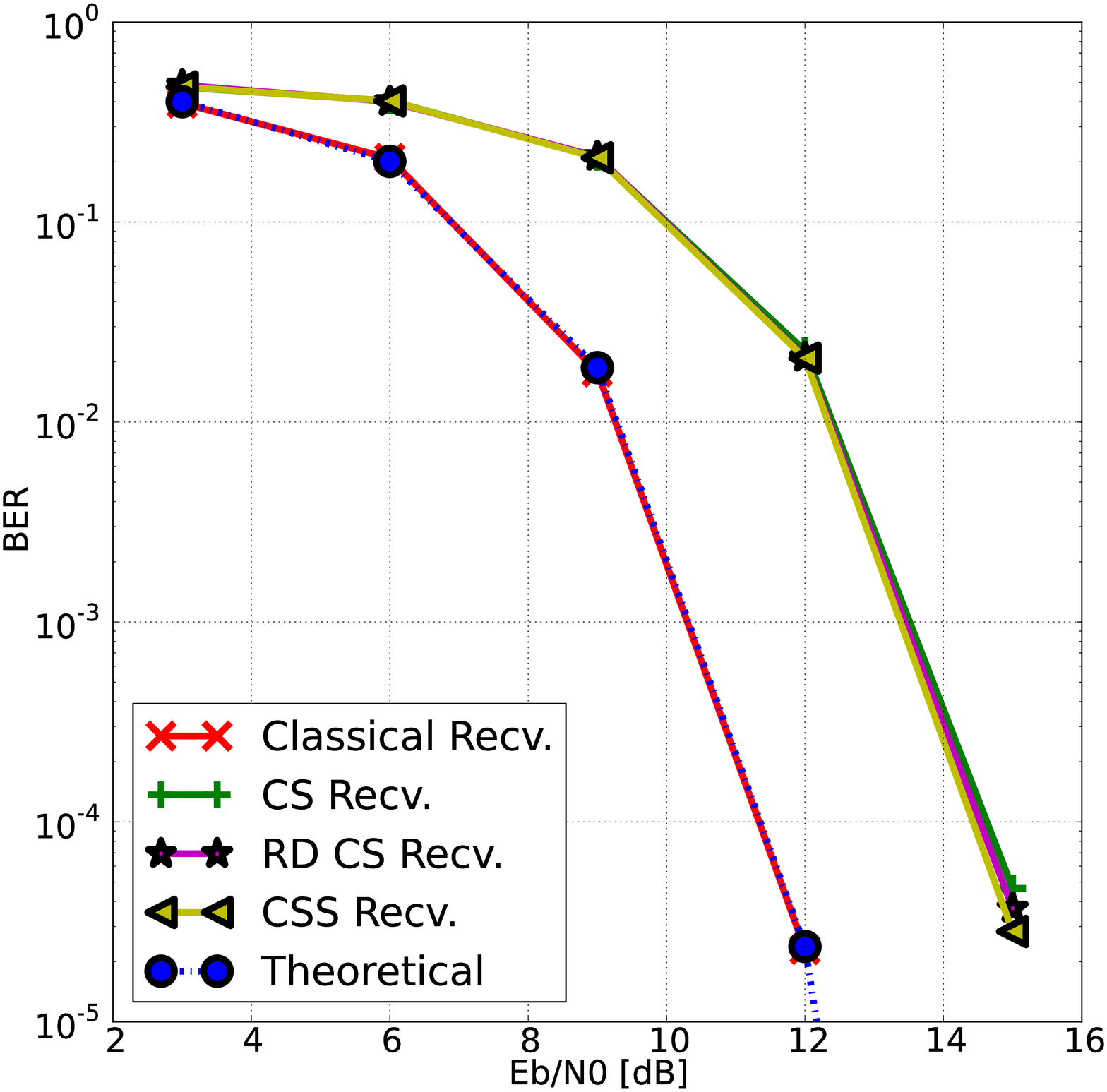}}
    \caption{BER versus Eb/N0 for different dictionary sizes. CS here is the Rademacher measurement scheme. Simulations were run until $100$ bit errors were found for each Eb/N0 point.}
    \label{fig:rf_numerical_experiments}
\end{figure*}

The results of the experiment are shown in \figref{fig:rf_numerical_experiments}.
The theoretical curve is calculated using a modified version of the non-coherent MFSK equation used before:
\begin{align}
    \nonumber P_b = \frac{M}{2(M-1)} \frac{1}{M} \sum_{k=2}^M (-1)^k {M \choose k} \cdot \\\exp\left(\text{log}_2(4)\frac{E_b}{N_0}\left(\frac{1}{k}-1\right)\right),
\end{align}
where $E_b/N_0$ is the energy per bit per noise spectral density and we multiply $E_b/N_0$ with $\log_2(4)$ because there are $4$ constellation points in QPSK.
As can be seen, the results here are close to identical with those for the simpler discrete numerical experiment.
Noise folding still gives rise to a penalty, which makes \CS a trade-off between sampling rate and \BER performance.
However, previous work has suggested that quantization may shift the trade-off point, so that \CS obtains both the low sampling rate and a better performance than a classical receiver \cite{Treichler2011}.
We investigate this in the following.

%%%%%%%%%%%%%%%%%%%%%%%%%%%%%%%%
\subsection{RF Numerical Experiment with Quantization}
\label{subsec:with_quantization}
\begin{figure}
    \centering
    \includegraphics[width=0.45\textwidth]{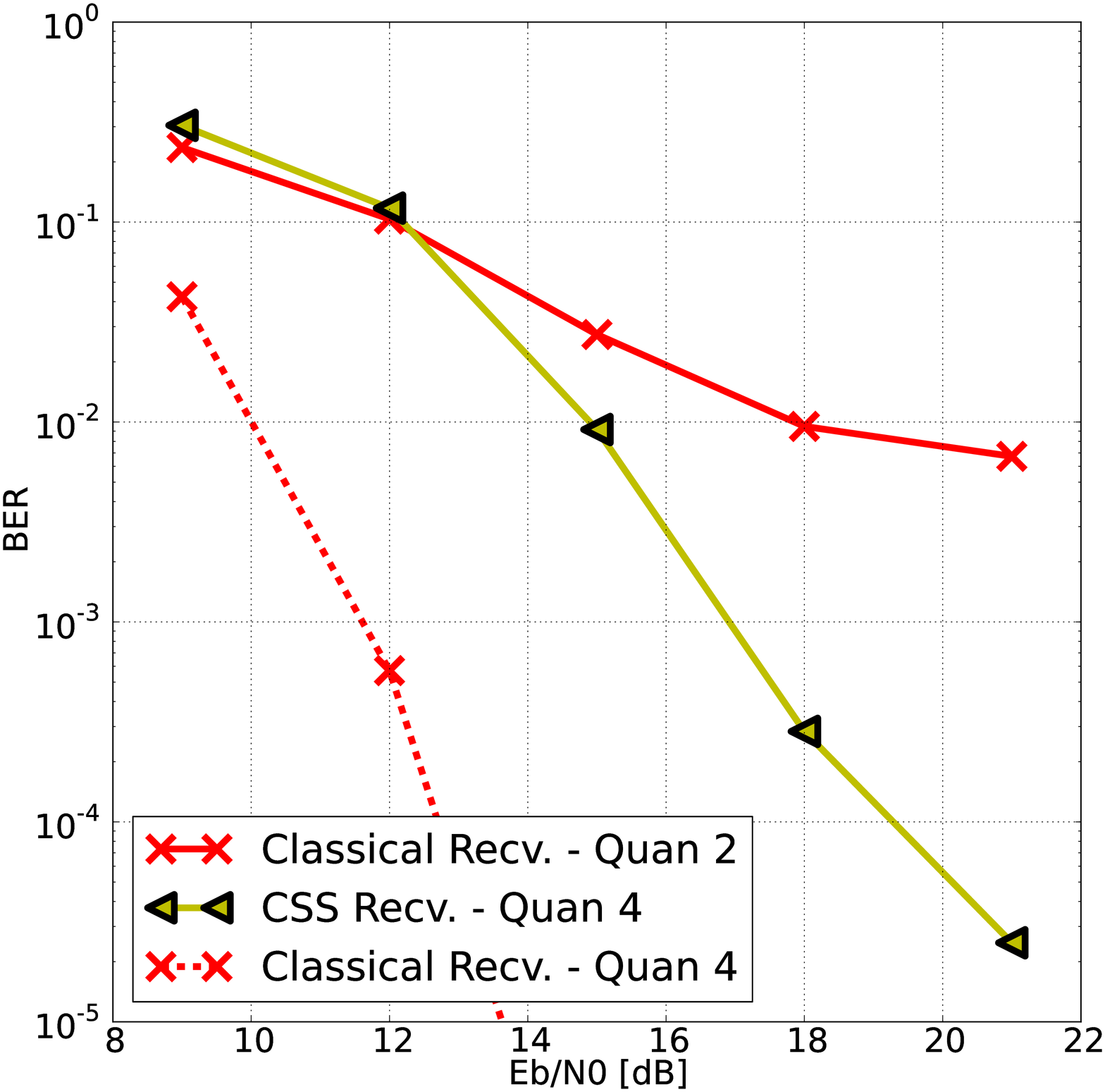}
    \caption{BER versus Eb/N0 for a classical receiver and a CSS receiver, both with quantization. $m=7$, $S=10$, $\kappa=2$ and $100$ errors found for each $E_b/N_0$ point.
    The dotted line is for a classical receiver with $4$ bits of quantization per sample.}
    \label{fig:quan_rf_m7_kappa2}
\end{figure}

%%%%%%%%%%%%%%%%%%%%%%%%%%%%%%%%
    In \cite{Treichler2011}, it is proposed to combat noise folding with quantization as a \CS receiver is able to quantize the sampled signal better, since it takes fewer measurements.
By better quantization we mean that if the \CS receiver takes half as many samples, it may quantize twice as well at no additional cost.
We have investigated this by applying uniform quantization to the RF experiment performed in the previous section.
However, as simple QPSK modulation is used, only the sign matters for demodulation and therefore quantization has no effect in the simple case of $S=1$ used so far. 
Therefore, we investigate $S=10$ instead and used $2$ bits of quantization per sample (i.e. $4$ bits of quantization for \CSS as $\kappa=2$).
This is merely intended as an example study to show that when taking into account quantization, \CS may perform better than a classical receiver.
The result of the numerical experiment is shown in \figref{fig:quan_rf_m7_kappa2}.
As can be seen, quantization makes \CS a better alternative in this scenario.
The \CS approach becomes significantly better for high $E_b/N_0$ values, because the classical receiver is not able to quantize the signal properly.
For comparison, we have also included the same result for a classical receiver with $4$ bits of quantization, i.e. the same level of quantization as the \CSS receiver.
Then it becomes clear that the classical receiver again is the best choice, but remember that it operates at twice the sampling frequency.
A CS-enabled receiver can therefore be seen as a trade-off point between sampling rate and dynamic range.

%%%%%%%%%%%%%%%%%%%%%%%%%%%%%%%
\section{Complexity Analysis}
\label{sec:complexity_analysis}
%%%%%%%%%%%%%%%%%%%%%%%%%%%%%%%
To evaluate the Subspace Pursuit algorithm, we investigate the computational complexity of the algorithm, shown in \tabref{tab:subpur_comp_cost},
where $K$ is the number of iterations used in the Subspace Pursuit algorithm, $S$ is the sparsity, $M$ is the number of measurements taken and $N$ is the number of Nyquist samples.
\begin{table*}[ht!]
\caption{Computational cost of the Subspace Pursuit algorithm.}
\label{tab:subpur_comp_cost}
\centering
\begin{tabular}{|l|l|}\hline
    \textbf{Action} & \textbf{Approx. cost} \\\hline
    \emph{Initialization:} & \\\hline
    $\bullet\hskip0.1cm 1$ computation of $\mA^\mathrm{T}\vy$ & $4MN$ \\\hline
    $\bullet\hskip0.1cm 1$ computation of $\vy-\mA_{T^0}\mA_{T_0}^\mathrm{T}\vy$ & $2M + 8MS$ \\\hline
    \emph{Loop:} & \\\hline
    $\bullet\hskip0.1cm K$ computations of $\mA^T\vy$ & $4KMN$ \\\hline
    $\bullet\hskip0.1cm K$ least squares problems ($\mA_{\tilde{T}}^\dagger\vy$) & $K(2M(2S)^2 + 11(2S)^3)$\\
    $\bullet\hskip0.1cm K$ computations of $\vy-\mA_{T^0}\mA_{T_0}^\dagger\vy$ & $K(2M + 4MS + 2MS^2 + 11S^3)$ \\\hline
    \textbf{Total:} & $99KS^3 + 4(K+1)MN$ \\
        & $+ 2(K+1)M + 4(K+2)MS + 10MKS^2$ \\\hline
\end{tabular}
\end{table*}

The matrix $\mA$ is real, but since $\vy$ is complex this affects the matrix-vector computations.
A matrix-vector product then costs $4MN$ and calculating a residual costs $2M + 8MS$.

The pseudo-inverse is never calculated, instead a linear least-squares problem is solved using the Singular Value Decomposition (SVD).
Solving a least-squares problem with $S$ variables and $M$ observations using the SVD costs \cite{Trefethen1997}:
\begin{align}
    \text{Cost}_{\text{LS with SVD}} \sim 2MS^2 + 11S^3.
\end{align}
Notice that the first least square problem in the loop takes in $2S$ atoms from the dictionary.
The cost of sorting and locating entries is not taken into account here, as those algorithms are more memory then computationally demanding.

It is also important to notice that the problem sizes involved here are very small.
Compressive sensing only works for sparse signals, so $S$ is often small compared to $M$ and $N$.
In the examples given here, $N=1023$ is the largest dimension we have worked with.
Because of this, the mathematical model in \tabref{tab:subpur_comp_cost} is not adequate, 
as the computational complexity is instead dominated by programming language overhead, such as the cost of calling different functions.
Therefore, it is important to include an extra term: $cK$, where $K$ is the number of iterations performed and $c$ is some constant that depend on system and programming language overhead.

\begin{figure}[b!]
    \centering
    \includegraphics[width=0.5\textwidth]{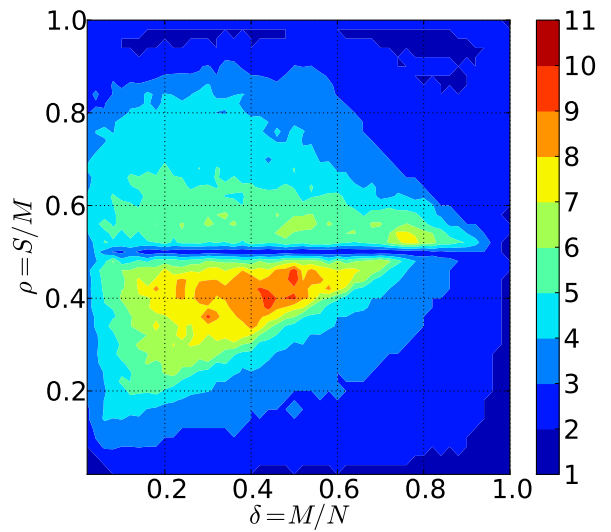}
    \caption{Number of Subspace Pursuit iterations for the CSS measurement matrix and Gold Dictionary size $m=10$.}
    \label{fig:subpur_iterations}
\end{figure}
It is of interest to investigate the required number of iterations, $K$, of the Subspace Pursuit algorithm, to better understand the cost of using \CSS.
In \figref{fig:subpur_iterations} we show the number of iterations used to generate the results in \figref{fig:phase_transitions}.
\begin{figure*}[ht!]
    \centering
    \includegraphics[width=1.0\textwidth]{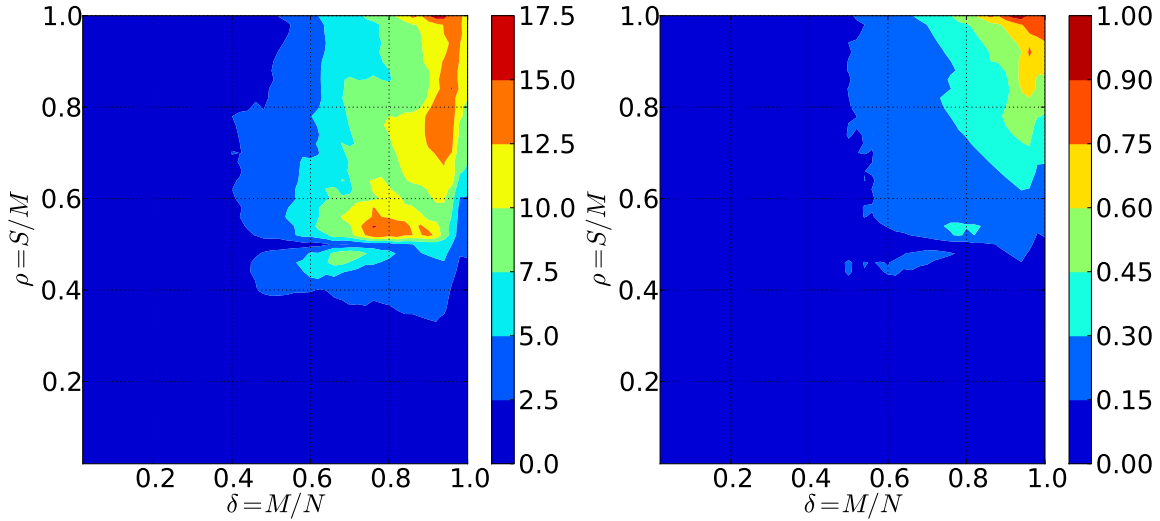}
    \caption{Computational cost of the Subspace Pursuit algorithm for the CSS measurement matrix and Gold Dictionary size $m=10$.
    The figure to the left shows measured execution time in seconds from the numerical experiment conducted in \figref{fig:phase_transitions} and the figure to the right is generated using the formula for the total computational cost found in \tabref{tab:subpur_comp_cost} plus the term $cK$ with $c=3\cdot 10^9$.
    For the figure to the right, the number of iterations of the Subspace Pursuit algorithm, $K$, is taken from \figref{fig:subpur_iterations} and the numbers are normalized.}
    \label{fig:subpur_compcosts}
\end{figure*}
The horizontal line through $\rho=0.5$ is interesting and unexpected.
If we change the input sparsity to the Subspace Pursuit algorithm from $S$ to $2S$, the line moves from $\rho=0.5$ to $\rho=0.25$, which means it is related to the number of atoms available to the Subspace Pursuit algorithm in each iteration.
It is not related to the dictionary type, as we have obtained exactly the same phase transition diagrams and iteration counts with a Haar wavelet packet dictionary.
Furthermore, it is not due to a "lucky" initial guess, as the line first emerges in the third iteration of the algorithm.
It seems to be an overlooked property of the algorithm, which has gone unnoticed so far because the line in \figref{fig:subpur_iterations} lies in the region of \figref{fig:phase_transitions},
where the algorithm cannot find the correct solution anyway.

Finally, we have measured the computation time for running the Subspace Pursuit algorithm for the \CSS numerical experiment in \figref{fig:phase_transitions}.
These are compared to the theoretical values obtained by using \tabref{tab:subpur_comp_cost}.
The constant $c$ has been set to $3\cdot 10^9$, which is a value found to give a good accordance with the numerically found values.
It is important to note that this choice of $c$ is very much a function of the algorithm, problem size, programming language and the machine on which the experiment is conducted and should therefore not be seen as a general choice.
The result is shown in \figref{fig:subpur_compcosts}.
The values in the figure on the right are normalized to one, as they are completely dependent on machine power and are only shown here to visualize how much the computational requirements change with the parameters.
As can be seen, the numerically obtained computation times seem to correspond fairly well to the mathematical model.
Each point in the above numerical experiment has been run as a simulation on $1$ out of $16$ threads on computation nodes with 2x Intel Xeon X5570 CPUs and 48GB memory.

%%%%%%%%%%%%%%%%%%%%%%
\section{Conclusion}
\label{sec:conclusion}
%%%%%%%%%%%%%%%%%%%%%%
In this work we apply \CS to a general \CDMA system and we show that it is possible to use a very simple measurement scheme at the receiver side to enable subsampling of the \CDMA signal.
We show that the performance of the proposed receiver scheme is affected negatively in \BER performance, similar to other \CS schemes.
However, we also show that when taking quantization into account, the proposed receiver model performs better in our example than a classical receiver with the same quantized bit rate.
Finally, we investigate the complexity of the developed algorithms and compare the computational cost of the numerical experiments with the theoretically calculated computation cost.

Our work here has shown that \CS used in spread spectrum receivers allows for a simplified front-end compared to other state-of-the-art \CS sampling designs.
Furthermore, we have shown that the problem of noise folding may be remedied in some cases by using quantization.
Future work should investigate further which scenarios may benefit from \CS and also perform laboratory experiments with the \CSS receiver structure.
Furthermore, the premise of this work is that taking fewer samples conserves power. This must be validated through laboratory experiments and the power efficiency of the \CSS receiver structure should be better evaluated.

\bibliographystyle{IEEEtran}
% argument is your BibTeX string definitions and bibliography database(s)
\bibliography{my_bibtex}

% biography section
\begin{IEEEbiography}[{\includegraphics[width=1in,height=1.25in,clip,keepaspectratio]{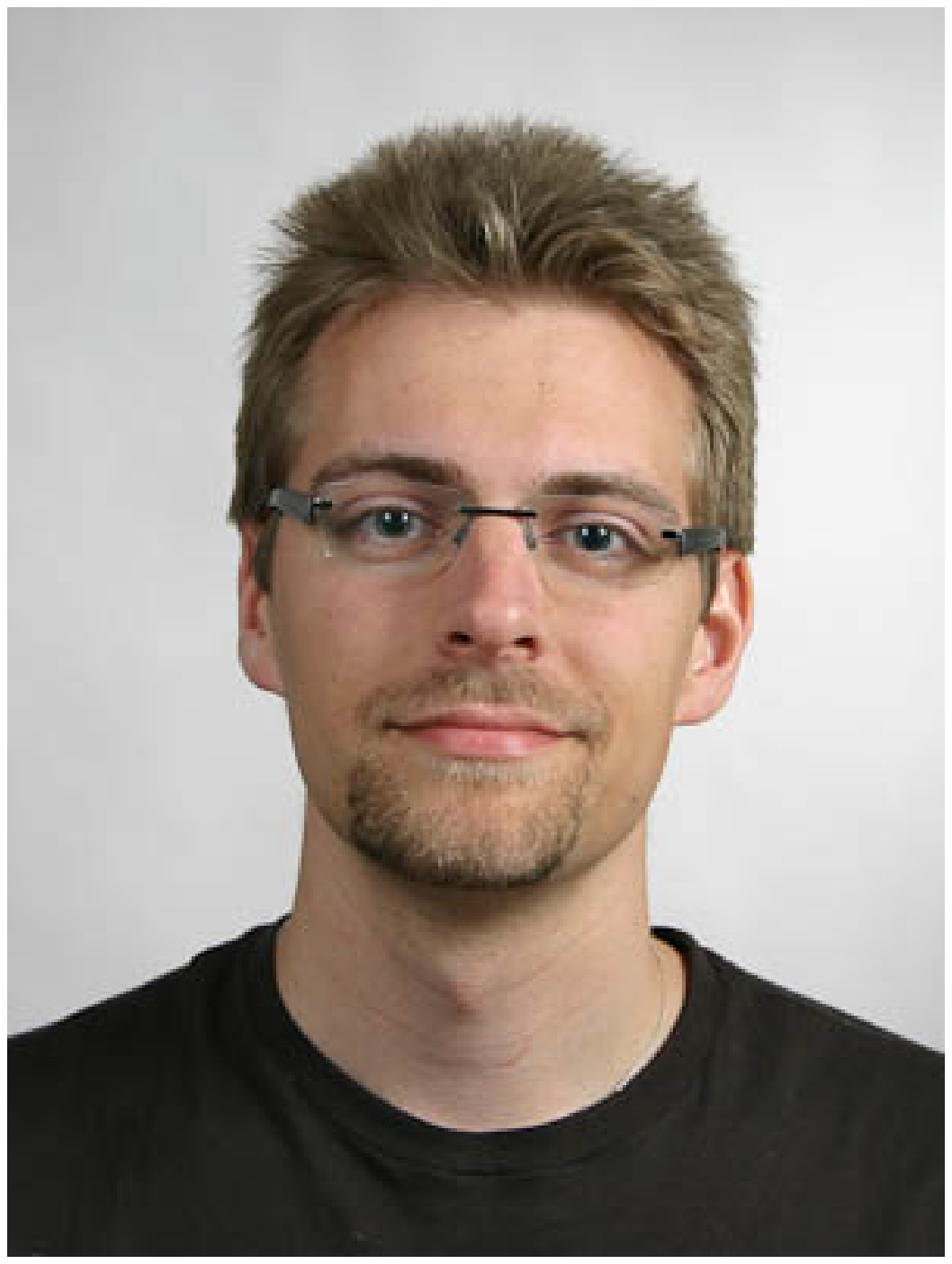}}]{Karsten Fyhn}
obtained his M.Sc. in wireless communication from Aalborg University, Denmark in 2010 as a student at the ELITE-programme at Aalborg University, 
a special research programme for highly skilled and motivated students.
He is currently a Ph.D. student at Aalborg University, working on compressive sensing in wireless communication.
In 2009 he was a research assistant at University of California Davis and in 2012 at Universiy of Massachusetts Amherst.
He is the recipient of the Elite Research Scholarship awarded by The Danish Ministry of Research, Innovation and Higher Education in 2012.
His interests include signal processing, compressive sensing, parameter estimation, manifold models, multi-packet reception, power-efficient wireless communication systems and wireless communication receiver structures.
\end{IEEEbiography}
\begin{IEEEbiography}[{\includegraphics[width=1in,height=1.25in,clip,keepaspectratio]{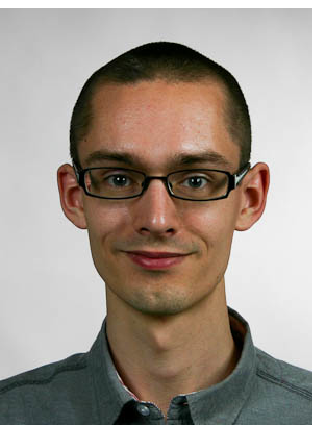}}]{Tobias Lindstrøm Jensen}
Tobias Lindstr{\o}m Jensen received the M.Sc. degree and Ph.D. degree in Electrical Engineering from Aalborg University in 2007 and 2011, respectively. 
Tobias L. Jensen is currently affiliated with the Department of Electronic Systems at Aalborg University.  
In 2007, he was an intern at Wipro-NewLogic Technologies in Sophia-Antipolis, France. 
In 2009 he was a visiting research scholar at University of California, Los Angeles (UCLA). 
His research interests includes optimization, signal and image processing, signal processing for communication, inverse problems and estimation.
\end{IEEEbiography}
\begin{IEEEbiography}[{\includegraphics[width=1in,height=1.25in,clip,keepaspectratio]{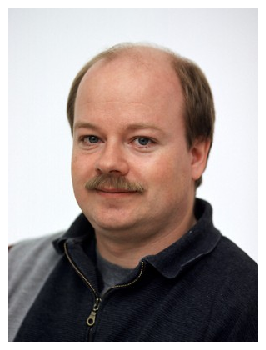}}]{Torben Larsen}
Torben Larsen (S’88, M’99, SM’04) received the M.Sc.E.E. and Dr.Techn. degrees from Aalborg University, Aalborg, Denmark, in 1988 and 1998, respectively. 
Since 2001, he has been a Full Professor at Aalborg University in electronic circuits, signals and systems theory. 
He has industrial experience working as senior engineer at Bosch Telecom and Siemens Mobile Phones. 
He was member in 2005-2010 and vice-chairman in 2009-2010 of the Danish Research Council for Technology and Production Sciences. 
In 2011 he was appointed director of the doctoral school at The Faculty of Engineering and Science, Aalborg University, with more than 650 enrolled PhD students. 
He has authored or co-authored over 130 peer-reviewed journal and conference papers and contributed to four internationally published books. 
He received the Spar Nord Research Prize in 1999, “Teacher of the year 2012” of the study board for electronics and IT at Aalborg University and 
later also the similar award from the Faculty of Engineering and Science. 
In the period 2012-2016 he is the Danish technical delegate of ESA’s Joint Board on Communications Satellite Program. 
Since 2007 he has been member of the Academy of Technical Sciences, Denmark, and IEEE Senior Member since 2004. 
In 2013 he was elected secretary of the IEEE Denmark section. He has supervised close to 20 PhD students. 
His recent research interests mainly include scientific computing, compressive sensing, numerical algorithms etc. in the areas of signals and systems theory.
\end{IEEEbiography}
\begin{IEEEbiography}[{\includegraphics[width=1in,height=1.25in,clip,keepaspectratio]{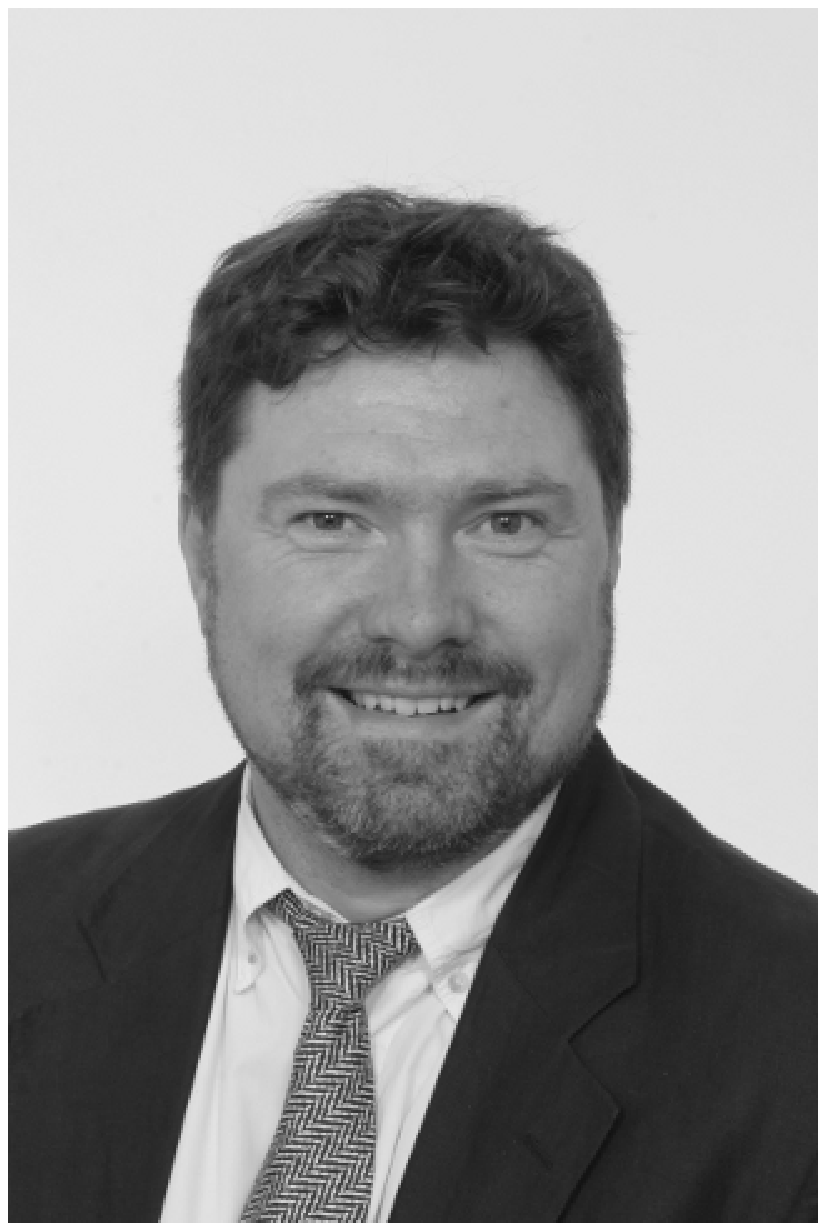}}]{S{\o}ren Holdt Jensen}
S{\o}ren Holdt Jensen (S’87–M’88–SM’00) received the M.Sc.\ degree in electrical engineering from Aalborg University, Aalborg, Denmark, in 1988, 
and the Ph.D.\ degree in signal processing from the Technical University of Denmark, Lyngby, Denmark, in 1995. 
Before joining the Department of Electronic Systems of Aalborg University, he was with the Telecommunications Laboratory of Telecom Denmark, Ltd, Copenhagen, Denmark; 
the Electronics Institute of the Technical University of Denmark; the Scientific Computing Group of Danish Computing Center for Research and Education (UNI$\bullet$ C), Lyngby; 
the Electrical Engineering Department of Katholieke Universiteit Leuven, Leuven, Belgium; and the Center for PersonKommunikation (CPK) of Aalborg University.
He is Full Professor and is currently heading a research team working in the area of numerical algorithms, optimization, 
and statistical signal processing for speech and audio processing, image and video processing, multimedia technologies, and digital communications.
Prof.\ Jensen was an Associate Editor for the IEEE Transactions on Signal Processing and Elsevier Signal Processing, 
and is currently Associate Editor for the IEEE Transactions on Audio, Speech and Language Processing. 
He is a recipient of an European Community Marie Curie Fellowship, former Chairman of the IEEE Denmark Section and the IEEE Denmark Section’s Signal Processing Chapter. 
He is member of the Danish Academy of Technical Sciences and was in January 2011 appointed as member of the 
Danish Council for Independent Research---Technology and Production Sciences by the Danish Minister for Science, Technology and Innovation.
\end{IEEEbiography}

\end{document}